\documentclass[twocolumn,aps,prd,showpacs,nofootinbib,superscriptaddress,floatfix,showkeys]{revtex4-2}

\usepackage{amsfonts}
\usepackage{amsmath}
\usepackage{amssymb}
\usepackage{bm}
\usepackage{dcolumn}
\usepackage{graphicx}   
\usepackage[latin1]{inputenc}
\usepackage{latexsym}
\usepackage{rotating}
\usepackage{hyperref}
\usepackage{graphicx}
\usepackage{color}

\newcommand\be{\begin{equation}}
\newcommand\ba{\begin{eqnarray}}
\newcommand\ee{\end{equation}}
\newcommand\ea{\end{eqnarray}}

\begin{document}

\title{ALP Dark Matter,  Cosmological Magnetic Fields and the Direct Collapse Black Hole Formation Scenario}

\author{Ashu Kushwaha}
\email{kushwaha.a.celb@m.isct.ac.jp}
\affiliation{Department of Physics, Institute of Science Tokyo,
2-12-1 Ookayama, Meguro-ku, Tokyo 152-8551, Japan}
  
\author{Robert Brandenberger}
\email{rhb@physics.mcgill.ca}
\affiliation{Department of Physics, McGill University, Montr\'{e}al,
  QC, H3A 2T8, Canada}
\affiliation{Trottier Space Institute, Department of Physics, McGill
University, Montr\'{e}al, QC, H3A 2T8, Canada}


\begin{abstract}
	
	Assuming that dark matter is an ultralight pseudoscalar particle which couples to electromagnetism like an axion (an ALP), we demonstrate that the coupling of the cosmological magnetic field produced by the ALP field oscillations to the primordial dark matter fluctuations yields a spectrum of gauge field fluctuations which can produce a sufficient flux of Lyman-Werner photons to enable the Direct Collapse Black Hole formation scenario.  The induced flux is consistent with the bounds on the excess flux of radio photons from ARCADE2 and EDGES measurements.

\end{abstract}

\maketitle

\section{Introduction} 
\label{sec:intro}

It has recently \cite{BFJ} \footnote{For earlier work see e.g. \cite{earlier}.} been realized that ALP dark matter \footnote{See e.g. \cite{Elisa} for reviews on ALP dark matter.}coherently oscillating on cosmological scales can produce cosmological magnetic fields of sufficient strength to explain observations \footnote{See e.g. \cite{Durrer} for a review of cosmological magnetic fields, and \cite{Vovk} for the observational results indicating that there is a lower bound on the strength of magnetic fields on cosmological scales.}. The mechanism is based on the Lagrangian
\be \label{Lag}
{\cal{L}} \, = \, \frac{1}{2} \partial_{\mu} \phi \partial^{\mu} \phi - \frac{1}{2} m^2 \phi^2 - g_{\phi, \gamma} F \wedge F \, ,
\ee
which describes the coupling of the axion-like particle (ALP) field $\phi$ to the field strength tensor $F$ of electromagnetism.  Here, $g_{\phi \gamma}$ is the constant coupling and has inverse mass dimensions.  If $\phi$ is oscillating coherently on cosmological scales \footnote{Such oscillations can be induced by a misalignment mechanism, e.g. the one suggested in \cite{misalign}.}, then the above coupling induces a tachyonic resonance for gauge field fluctuation modes with wavenumber $k < k_c$,  where $k_c$ is the critical wave number whose value is given below.  The instability to gauge field production sets in right after the time of recombination when plasma effects become negligible.

As pointed out in \cite{Nirmalya}, there is also a narrow parametric resonance instability which for certain field values dominates. A similar mechanism for magnetogenesis also exists if the ultralight dark matter field is a scalar field which couples to $F^2$ \cite{Vahid}. 

Another mystery in our current view of cosmology is the origin of super-massive black holes which are observed at high redshifts.  These black holes may be primordial (see e.g. \cite{Carr} for a review).  If they are not, then there are two main challenges. First, a sufficient number of nonlinear seed fluctuations on the relevant scales must be present, and secondly, matter collapsing onto the seed fluctuations must be prevented from fragmenting \footnote{Recall that it is harder to turn small objects into black holes than large objects.}.  A sufficient condition to prevent fragmentation is the presence of a sufficient flux of Lyman-Werner (LW) photons. In \cite{Jiao} it was pointed out that the photons created by the instability discussed in \cite{BFJ} may create such a flux, provided that there is a sufficiently efficient cascade of photons from infrared scales to the Lyman-Werner range. 

In this paper we point out that there is a direct channel to produce Lyman-Werner photons by coupling the magnetic field $B$ generated by the above-mentioned instability to the primordial density fluctuations which are represented at late times on sub-Hubble scales by fluctuations in the dark matter density and hence of the dark matter field.  Thus, instead of considering the process $\phi \rightarrow A_k + A_{-k}$ (where $A_k$ stands for the electromagnetic gauge field mode with wavelength $k$) we consider the process $\delta  \phi_k + B \rightarrow \delta A_k$, viewing the magnetic field as uniform. We compute the induced spectrum of $A_k$, compute the resulting flux of Lyman-Werner photons and show that this flux may be sufficiently high to satisfy the Direct Collapse Black Hole (DCBH) criteria. We show that the resulting flux is consistent with the limits which can be inferred by the ARCADE2 \cite{Arcade} and EDGES \cite{Edges} result, in both cases taking the claimed detections as upper bounds.

\section{Production of Secondary Photons}

Our starting point is the Lagrangian (\ref{Lag}) which describes the coupling of $\phi$ to the electromagnetic gauge field.  We consider the time evolution after the parametric instabilities have generated the cosmological magnetic field.  Since we are interested in length scales which are very small compared to the scale of the magnetic field $B$, we can take $B$ to be homogeneous.  We wish to compute the photons produced by the magnetic field $B$ interacting with the primordially-induced dark matter field fluctuations $\delta \phi$. 

As reviewed in the Appendix, the equation of motion which describes the evolution of the gauge field fluctuations $A_i$ (we are working in Coulomb gauge) in the presence of background magnetic and electric fields $B_i$ and $E_i$ is
\be \label{EoM1}
A_i^{\prime \prime} - \nabla^2 A_i \, = \, S_i \, ,
\ee
with the source $S_i(\eta)$ given by
\be 
S_i \, = \, - g_{\phi \gamma} \phi^{\prime} B_i + g_{\phi \gamma} \epsilon^{ijk} \partial_j \phi E_k \, .
\ee
In the above equation, a prime indicates the derivative with respect to conformal time $\eta$, and $\epsilon^{ijk}$ is the totally antisymmetric symbol.  

In the case of oscillating ultralight dark matter we have two free parameters, the mass $m$ of the dark matter field and the coupling constant $g_{\phi \gamma}$. We use the following parametrization
\be
m \, \equiv \, m_{20} 10^{-20} {\rm eV} \, ,
\ee
and
\be
g_{\phi \gamma} \, \equiv \, {\tilde{g}}_{\phi \gamma} 10^{-10} {\rm GeV}^{-1} \, .
\ee
Recent studies provide a bound of $m_{20} > 10$ if $\phi$ makes up all of the dark matter \cite{Dalal},  and a recent bound on the coupling constant range is $g_{\phi \gamma} < 10^{-9} m_{20} {\rm{GeV}}^{-1}$ \cite{bound} (see also \cite{bound2} and \cite{website} for a website with a summary of present and future constraints).
 
As shown in Ref.\cite{BFJ}, the tachyonic resonance leads to magnetic field generation on scales $k < k_c$ with
\be
k_c \, \simeq \, g_{\phi \gamma} m \phi_0 a(\eta) \, ,
\ee
where $\phi_0$ is the amplitude of the oscillation of the background $\phi$ field and $a(\eta) $ is the scale factor.  In fact, the magnetic field strength is highly peaked at this value of $k$. If $\phi$ makes up all of the dark matter at the time of recombination, then $m \phi_0 \sim T_{\rm rec}^2$ (where $T_{\rm rec}$ is the temperature at recombination) and the wavelength corresponding to $k_c$ becomes
\be
\lambda_c \, \sim \, {\tilde{g}}_{\phi \gamma}^{-1} 10^{-6} {\rm{Mpc}} \, ,
\ee
which is enormous compared to the length scales considered in this work.

As was found in \cite{Nirmalya} in the case when ${\tilde{g_{\phi \gamma}}} m_{20}^{-1} < 1$ the tachyonic resonance is not efficient,  and narrow band parametric resonance (as studied first in the context of inflationary reheating in \cite{TB, DK}) takes over.  The instability band is centered about the $k$ value 
\be
k \, \sim am \,\, \rightarrow \,\, \lambda(k) \, \sim \, m_{20}^{-1} 10^{-6} {\rm{Mpc}}
\ee
and the corresponding wavelength is thus also much larger than the length of relevance for the present study. Hence, it is well justified to consider the magnetic field to be uniform in this study.

To compute the secondary gauge fields, we will be working to linear order in the amplitude ${\cal{A}}$ of the cosmological perturbations and treat $B(\eta)$ as a background field (we take $E = 0$).  Thus, we Fourier expand (\ref{EoM1}) to obtain
\be \label{EoM2}
A_i(k)^{\prime \prime} + k^2 A_i(k) \, = \, - g_{\phi \gamma} (\delta \phi)^{\prime}(k) B_i (\eta) \, \equiv \, S_i(k, \eta) \, ,
\ee
where $(\delta \phi)(k)$ is the fluctuation of the dark matter field induced by the cosmological fluctuations. We are interested in secondary photons produced after recombination (when plasma effects become negligible) and hence consider vanishing initial fluctuations at the time $\eta_{\rm rec}$ of recombination \footnote{If there are pre-existing fluctuations, they would simply add to the contribution we are computing.}
\be \label{IC}
A_i(\eta_{\rm rec}) = A_i^{\prime}(\eta_{\rm rec}) = 0 \, .
\ee
The equation (\ref{EoM2}) can be solved using the Green's function method. If $G(\eta, \tau)$ is the Green's function of the homogeneous part of (\ref{EoM2}) (which does not depend on $i$, the component of the gauge field mode) then the solution of the inhomogeneous equation given the vanishing initial conditions (\ref{IC}) is
\be \label{EoM3}
A_i (k, \eta) \, = \, \int_{\eta_{\rm rec}}^\eta G(k; \eta, \tau) S_i(k, \tau) d\tau \, .
\ee

The Green's function $G(k; \eta, \tau)$ is a solution of
\be
G(k)^{\prime \prime} + k^2 G(k) \, = \, \delta(\eta - \tau) 
\ee
and is given by the following well-known combination of the linearly independent solutions $f_1(k; \eta)$ and $f_2(k; \eta)$ of the homogeneous equation
\be
G(k; \eta, \tau) \, = \, \frac{1}{\epsilon(k; \tau)} \bigl( f_1(k; \eta) f_2(k; \tau) - f_2(k; \eta) f_1(k; \tau) \bigr) \, ,
\ee
with
\be
\epsilon(k; \tau) \, = \, f_1(k; \tau) f_2^\prime(k; \tau) - f_2(k; \tau) f_1^\prime(k; \tau) \, .
\ee
In the case of our simple homogeneous differential equation the functions $f_1$ and $f_2$ are simple oscillations and the Green's function becomes
\be \label{GF}
G(k; \eta, \tau) \, = \, \frac{1}{k} {\rm{sin}}(k(\eta - \tau)) \, 
\ee
for $\eta \geq \tau$, and vanishing for $\eta < \tau$.

\section{Secondary Photons from Cosmological Perturbations}

So far, the analysis was general.  Now we are interested in evaluating the flux of secondary photons which come from the coupling of a cosmological magnetic field of strength $B$ (taken to be homogeneous on the scales  relevant to us) with the cosmological fluctuations which on sub-Hubble scales are manifest as fluctuations in the dark matter field.  We take the spectrum of primordial cosmological perturbations to be scale-invariant.  This implies that the power spectrum of the density fluctuations is independent of the wavenumber when measured at the time $t_i(k)$ when the scale $k$ crosses the Hubble radius, i.e.
\be
\left(\frac{\delta \rho}{\rho}\right)^2(k, t_i(k)) k^3 \, = \, {\cal{A}} \, ,
\ee
where ${\cal{A}}$ is a constant. Its value is determined by the amplitude of the cosmic microwave background (CMB) fluctuations, ${\cal{A}} \sim 10^{-9}$.  Scales we are interested in enter the Hubble radius long before the time of recombination. But they are frozen (modulo logarithmic corrections) until the time of equal matter and radiation (see e.g. \cite{MFB, RHBrev} for reviews of the theory of cosmological fluctuations). Hence, the power spectrum is independent of $k$ on the scales of interest at the time of recombination.

Considering the potential of $\phi$, which can be approximated by 
\be
V(\phi) \, = \, \frac{1}{2} m^2 \phi^2 \, ,
\ee
the field fluctuation $\delta \phi(k)$ is given by
\be
\delta \phi(k) \, = \, \frac{1}{2} \phi_0 \frac{\delta \rho}{\rho}(k) \, = \, \frac{1}{2} \phi_0 k^{-3/2} {\cal{A}}^{1/2} \, .
\ee
In the above, $\phi_0$ is the amplitude of oscillations of $\phi$.  Under the assumption that $\phi$ makes up all of the dark matter at recombination, we have 
\be \label{DMrel}
m^2 \phi_0^2 \, \sim \, T_{rec}^4 \, .
\ee

Let us now estimate the spectrum of secondary photons, assuming that there are no such photons present at the time of recombination \footnote{If there were photons present, then the flux computed here would simply add linearly to the flux present initially.}. We use the value of the B-field (denoted by B) right after recombination and take into account the redshifting of this field $\sim a^{-2}(\tau)$ after that time. Inserting the expression (\ref{GF}) for the Green's function into the expression (\ref{EoM3}) with the source function given by the right hand side of (\ref{EoM2}) we obtain for the magnitude of the A-field
\ba
A(k, \eta) && \, \sim \, g_{\phi \gamma} B {\cal{A}}^{1/2} k^{-3/2} \phi_0 \times \\ 
\int_{\eta_{rec}}^{\eta} &d\tau& {\rm{sin}} k(\eta - \tau) {\rm{cos}} k(\tau - \eta_{rec})
\bigl( \frac{a(\eta_{rec})}{a(\tau)} \bigr)^{2} \, .
\nonumber
\ea
From the last factor it follows that the production of secondary photons is dominated by those produced close to the time of recombination.  By changing variables from $\tau$ to $x \equiv k\tau$ it follows immediately that the order of magnitude of the integral is $1/k$. 
Hence, we obtain the following expression for the energy density spectrum of secondary photons (at a time close to recombination):
\be
\rho_A(k) \, \sim \, k^2 A_k(\eta)^2 \, .
\ee
This yields a flux $\Phi_A(k)$ \footnote{Note that this is a number density and scales as $a(t)^{-3}$, the same way as the $\omega^2 \delta T(\omega)$ factor appearing in (\ref{ratio}).} of photons at wavenumber $k$ given by
\ba
\Phi_A(k) \, &\sim& \, \frac{1}{k} k^3 \rho_A(k) \nonumber \\
&\sim& \, g_{\phi \gamma}^2 B^2 {\cal{A}} k^{-1} \phi_0^2 \, .
\ea


Inserting the expression for $\phi_0$ which follows from (\ref{DMrel}), using the value $B \sim 1 {\rm{Gauss}}$ which is the amplitude of the magnetic field on the scale $k_c$ produced by the tachyonic resonance (or on the corresponding scale for narrow band parametric resonance),  and evaluating the flux for a value of $k\sim 10 \, {\rm  eV}$ in the Lyman-Werner band we obtain  
\be \label{flux}
\Phi_A(k) \, \sim \, {\tilde{g}}_{\phi \gamma}^2 m_{20}^{-2} 10^{-39} {\rm{GeV}}^3 \, .
\ee
Provided that
\be \label{result1}
{\tilde{g}}_{\phi \gamma}^2 m_{20}^{-2} \, > \, 10^{-5} \, ,
\ee
we obtain a flux of Lyman-Werner photons sufficiently large to prevent fragmentation of a collapsing neutral hydrogen cloud. Thus, the key criterion to open up the Direct Collapse Black Hole formation channel for super-massive black hole formation can be satisfied.

\section{Constraints from ARCADE2}

A flux of excess radio photons is claimed to have been seen by the ARCADE2\cite{Arcade} and EDGES \cite{Edges} experiments, although these results are not generally accepted as real detections, and the explanation of this excess remains an open question.  We can, however, use the results to set an upper bound on the flux of secondary photons on the relevant frequency scales.  We will use the ARCADE2 constraint 
\be \label{crit}
\frac{\delta T(\omega)}{T_{CMB}} \, < \, 4 \times 10^{-4}
\ee
at a frequency $\omega$ given by 
\be 
\frac{\omega}{T_{CMB} \, \sim \, 0.2 \, ,}
\ee
where $T_{CMB}$ is the background CMB temperature, and $\delta T(\omega)$ is the brightness temperature of the secondary photons at the frequency $\omega$

Making use of
\be \label{ratio}
\delta T(\omega) \, = \, \frac{4 \pi^2}{\omega^2} \Phi_A(\omega) 
\ee
and making use of the result (\ref{flux}) for the flux computed in the previous section we find that the condition (\ref{crit}) is obeyed provided that\footnote{Note that from EDGES \cite{Edges}, we have $\frac{\delta T(\omega)}{T_{CMB}} \, < \, 1$ for $\frac{\omega}{T_{CMB}} \, \sim \, 1.4\times 10^{-3}$, which gives the upper bound on the parameter space $
	{\tilde{g}}_{\phi \gamma}^2 m_{20}^{-2} \, < 5 \times \, 10^4 \,
	$.}
\be \label{result2}
{\tilde{g}}_{\phi \gamma}^2 m_{20}^{-2} \, < \, 4 \times 10^5 \, .
\ee
To obtain both results (\ref{result1}) and (\ref{result2}) we have used the value $B \sim 1 {\rm{Gauss}}$ for the strength of the magnetic field. If only a small fraction ${\cal{F}}$ of the dark matter density converts into the magnetic field $B$, then the results get modified.

Note that demanding that we get a sufficient flux of LW photons puts a lower bound on the parameter combinations ${\tilde{g}}_{\phi \gamma}^2 m_{20}^{-2} $, the ARCADE2 results give an upper bound.  So, in this sense, we have a parameter space bounded both from above and below.

\section{Discussion and Conclusions}

We have computed the usual ALP scenario in which an ultralight pseudoscalar field $\phi$ is coupled to the field strength of electromagnetism via a $\phi F \wedge F$ term in the Lagrangian.  We have considered the flux of secondary photons produced by a background magnetic field $B$ interacting with the dark matter fluctuations $\delta \phi$ which stem from the primordial cosmological fluctuations. We have shown that for a range of parameter values this mechanism is sufficiently strong to generate the Lyman-Werner background required to prevent the fragmentation of matter clouds collapsing onto primordial seed fluctuations. Hence, our mechanism provides a channel to enable the DCBH scenario of super-massive black hole formation. 

Our analysis is general. If $\phi$ is the dark matter field, then the fluctuations $\delta \phi$ which we have used are the ones which emerge from the observed cosmological fluctuations.  We do not need to assume a particular model for these fluctuations. They could be produced by primordial inflation,  but also by one of the alternatives (see e.g. \cite{RHBalt} for a discussion of alternatives to inflation, and \cite{TCC} for some recent challenges to effective field theory models of inflation). We do need to assume that the spectrum of fluctuations extends from the CMB scales all the way down to the small length scales corresponding to the value of $k = 10 {\rm{eV}}$ at the time of recombination.  Whether this is the case or not will depend on the specific mechanism producing cosmological perturbations in the very early universe.

The motivation of our work came from the realization \cite{BFJ} that oscillating ultralight dark matter coupled to electromagnetism via (\ref{Lag}) can generate magnetic fields on infrared scales.  The mechanism, however, works also if the magnetic field $B$ has a different origin.  With minor changes, our mechanism also works if the dark matter is a scalar field which couples to $F^2$, as studied in \cite{Vahid}.

We have also studied the constraints on the coupling constant $g_{\phi \gamma}$ which can be derived by demanding that the induced flux of radio photons does not overproduce the excess signals reported by ARCADE2 and EDGES (taking the reported values as upper bounds on the possible flux).  We find that the constraints are satisfied for parameter values of the pseudoscalar mass which are consistent with other observational constraints.

There are a lot of avenues for further research. For example,  in this work we have treated the magnetic field as constant on the scales relevant to the constraints we are studying. It would be interesting to generalize the work and include the effects of the possible presence of initial magnetic field fluctuations on smaller scales. In this case, the source term in (\ref{EoM1}) would become a convolution integral coupling $\delta \phi(k^\prime)$ to $B(k - k^\prime)$. 

In this work we have taken the spectrum of cosmological perturbations to be scale-invariant, in agreement with observations on cosmological scales (we have neglected the effects of the small red tilt). However, it is possible that the spectrum could be enhanced on smaller scales. In fact,  such an enhancement is often postulated in scenarios of primordial black hole formation.  Based on our study, a large enhancement of the spectrum on scales relevant to ARCADE2 and EDGES would be in conflict with observations. It would also be interesting to work this out in detail.

\section*{Acknowledgement}

\noindent 
We wish to thank the CERN Theory Group and in particular Valerie Domcke for inviting us to CERN where this collaboration started. The work of A.K. was supported by the Japan Society for the Promotion of Science (JSPS) as part of the JSPS Postdoctoral Program (Grant Number: 25KF0107). 
The research at McGill is supported in part by funds from NSERC and from the Canada Research Chair program.    

\section*{Appendix}

In this Appendix we derive the equation of motion (\ref{EoM1}) which is the basis of this work.  We consider electrodynamics in a flat FRW universe whose metric in conformal time ($\eta$) is given by
\begin{align}\label{metric}
	ds^2 =  a^2(\eta) \left[ d\eta^2 - \delta_{ij} dx^i dx^j \right]~~.
\end{align}
where $a(\eta)$ is the scale factor in conformal time, $\eta$. The four velocity of the comoving observer is given as $u^{\mu} = (a^{-1}(\eta),0,0,0)$ and $u_{\mu} = (a (\eta),0,0,0)$ with $u_\mu u^\mu = 1$. The electric and magnetic fields are defined as $E_\mu = F_{\mu\nu} u^\nu$ and $B_\mu = \frac{1}{2} \epsilon_{\mu\nu\alpha\beta} u^\nu F^{\alpha\beta}$, which in Coulomb gauge ($A^0 = 0, \partial_i A^i = 0$) gives
\begin{align}
	E_i &= -\frac{1}{a} F_{0i} = -\frac{1}{a} A_i' \nonumber\\
	B^i &= \frac{1}{2 \, a^3}\eta^{i0jk} F_{jk} = \frac{1}{2 \, a^3}\varepsilon^{ijk} F_{jk} = \frac{1}{a^3}\varepsilon^{ijk} \partial_j A_k~~.
\end{align}
Therefore, we obtain the following 
\begin{align}
	E_{\mu} &= \left(0, -\frac{1}{a} A_i' \right) = a (0,\mathcal{E}),
	\nonumber\\ B_{\mu} &= \left(0, -\frac{1}{a} \varepsilon_{ijk} \partial_j A_k \right) = a (0,\mathcal{B})
\end{align}
where $\mathcal{E}, \mathcal{B}$ are Euclidean three vector field as decay as $a^{-2}$ in highly conducting plasma.  Now, we can define the rescaled fields $\textbf{E} \equiv a^2 \mathcal{E}$ and $\textbf{B} \equiv  a^2 \mathcal{B}$, which gives the following useful relations 
\begin{align}\label{rescaled-EB}
	{E}_i = -A_i' ~, \qquad {B}_i = -\varepsilon_{ijk} \partial_j A_k~~.
\end{align}
One advantage of working with the above rescaled fields is that we can obtain the equations of motion in terms of gauge fields and write them in terms of the rescaled fields, which takes the same form as Maxwell's equations in Minkowski spacetime.
For example, the energy density of the magnetic field follows the relation $\rho_B \equiv \frac{1}{2} a^{-4} \textbf{B}\cdot \textbf{B} = \frac{1}{2} \mathcal{B}\cdot \mathcal{B}$.

The dynamics of the axion and the EM field are determined by the Klein-Gordon and modified Maxwell equations as
\begin{subequations}\label{kg-maxwell-eqns}
	\begin{align}\label{kg-eq}
		\frac{1}{\sqrt{-g}}\partial_{\mu} \left( \sqrt{-g}\partial^{\mu} \phi \right) + m^2 \phi &=  \frac{1}{4}g_{\phi\gamma} F_{\mu\nu} \tilde{F}^{\mu\nu}  ~~, \\
		\label{maxwell-eq}
		\partial_{\mu}  \left( \sqrt{-g} \, F^{\mu\nu} \right) &= g_{\phi\gamma} \partial_{\mu} \left( \sqrt{-g} \phi \, \tilde{F}^{\mu\nu}  ~ \right)  
	\end{align}
\end{subequations}

Since we aim to study the generation of secondary photons due to the fluctuations in the scalar field $\delta \phi (\eta,\textbf{x})$, we can decompose the EM field tensor as $F_{\mu\nu} = \bar{F}_{\mu\nu} + \delta F_{\mu\nu}$, where 
$\bar{F}_{\mu\nu}$ refers to the background part corresponding to the cosmological magnetic field generated by the oscillating $\bar{\phi}$ after recombination. The produced secondary photons are determined by the EM field perturbation $\delta F_{\mu\nu}$ which refers to the propagating EM waves (radiation).
From Eq.~\eqref{kg-maxwell-eqns} in Coulomb gauge 
, we obtain
\begin{subequations}\label{kg-maxwell-eqns-1}
	\begin{align}\label{kg-eq-1}
		\phi'' + 2\mathcal{H} \phi' - \delta^{ij} \partial_i \partial_j\phi + m^2 a^2 \phi & = \frac{1}{4} a^2 g_{\phi\gamma} F_{\mu\nu} \tilde{F}^{\mu\nu} \\
		\label{maxwell-eq-1}
		\delta^{ij}\partial_0 F_{0j} - \delta^{lj} \delta^{ik} \partial_l F_{jk} &= \frac{1}{2} g_{\phi\gamma}\varepsilon^{ijk} \partial_0 (\phi F_{jk}) \nonumber\\ 
		&\quad- g_{\phi\gamma}\varepsilon^{ijk} \partial_j (\phi F_{0k})
	\end{align}
\end{subequations}
The secondary photons are denoted by the gauge field $\textbf{A}$ as $\delta F_{ij} = \partial_i A_j - \partial_j A_i$. From the above equation, we obtain Maxwell equation
\begin{align}\label{maxwell-eq-2}
	A_i'' - \nabla^2A_i = g_{\phi\gamma}\varepsilon^{ijk} \delta \phi' \partial_j \bar{A}_k - g_{\phi\gamma}\varepsilon^{ijk} \partial_j \delta \phi \, \bar{A}_k'
\end{align}
Now, using the definition of rescaled field for the background electric and magnetic field part ${E}_i = -\bar{A}_i' \, , ~~{B}_i = -\varepsilon_{ijk} \partial_j \bar{A}_k$, we obtain
\begin{align}\label{maxwell-eq-3}
	A_i'' - \nabla^2A_i = -g_{\phi\gamma}\delta \phi' {B}_i + g_{\phi\gamma}\varepsilon^{ijk} \partial_j \delta \phi \, {E}_k
\end{align}
This equation describes the production/evolution of secondary photons (left hand side of the equation) which are sourced by the interaction of scalar field fluctuations and background electric and magnetic field (right hand side).

\end{document}